\documentstyle[12pt,here,epsf,a41,rotate]{article}

\begin{document}

\begin{center}
{\LARGE\bf {\boldmath $\Delta G(x)$} from high {\boldmath $p_t$} hadrons in DIS at
a polarised HERA}

\vspace{1cm}
{G. R\"adel$^a$ and A. De Roeck$^{b}$}

\vspace*{1cm}
{\it $^a$CERN - Div.\ PPE, 1211 Gen\`eve 23, Switzerland.}\\
\vspace*{3mm}
{\it $^b$DESY - FH1, Notkestr.\ 85, 22603 Hamburg, Germany}\\

\vspace*{2cm}

\end{center}

\begin{abstract}
We investigate the possibility to identify  photon-gluon fusion (PGF)  events
in polarised deep inelastic $ep$
scattering, assuming the kinematics of the HERA collider,
by a pair of charged high $p_t$ particles.
In a Monte Carlo study we find possible selection criteria and
show the expected measurable asymmetries. We discuss the sensitivity
to $\Delta G(x)$ and compare the  result to the one obtained
using di-jets to tag PGF events.
\end{abstract}

\section{Introduction}

\vspace{1mm}
\noindent

The extraction of the polarized gluon density $\Delta G$ is one of the 
future key measurements in polarized physics. 
For  the polarized HERA collider several methods have been 
proposed to measure $\Delta G$~\cite{workshop}, 
the most promising one being the 
measurement of the di-jet asymmetry in deep inelastic scattering
events~\cite{gaby}.
In this paper we present a new method to access the polarized gluon
density at HERA, using two charged particles with 
high transverse momentum $p_t$, instead of jets. This 
analysis follows closely the one proposed for the COMPASS~\cite{compass} 
experiment at CERN, detailed in ref.~\cite{bravar}. The technique
to extract the (unpolarized) gluon density from high 
$p_t$ hadrons  has 
already been successfully applied
at HERA, albeit based on one single high momentum
charged track and for photoproduction data~\cite{H1gamma}.

\begin{figure}[t]
\hfil \epsffile{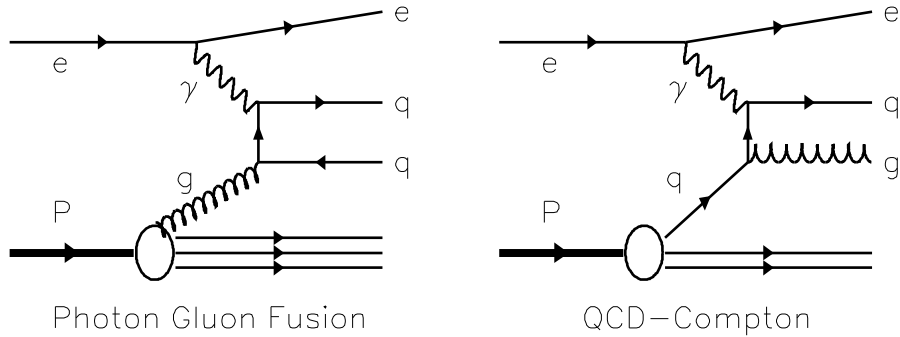} \hfil
 \hspace*{12cm}
\caption{Feynman diagrams for the di-jet cross sections at LO,
the Photon-Gluon Fusion (PGF) process (left) and the QCD-Compton 
process (QCDC) 
(right). }
\label{LO}
\end{figure}

The method is very similar to the one using  di-jet events to extract
$\Delta G$. The aim is to isolate subprocesses which are 
sensitive to the gluon at the Born level. Fig.~\ref{LO} shows that  
this is the case for the Leading Order (LO) photon gluon fusion process
(PGF).
In case of high partonic $p_t$ the two quarks will emerge as  
observable jets. These jets can be detected either calorimetrically,
as exploited in \cite{gaby} or can be tagged via particles with 
high transverse momentum with respect to the $\gamma^* p$ axis 
in the $\gamma^* p$ centre of mass system
(CMS), as studied here. The present experiments at the HERA collider
are equipped with excellent tracking detectors in the photon fragmentation
and central region, where most of the jets are expected. Hence we will 
use charged particles (tracks) for the  high $p_t$ 
hadrons to tag the PGF process.

In this exploratory study for HERA we are  guided by the 
analysis 
of the di-jet events, and impose similar criteria to select the PGF
process. Since in LO the two partons  are produced
with opposite $p_t$, two 
charged hadrons with approximately opposite azimuthal angles  $\phi$ 
in the $\gamma^* p$ CMS will 
be required. The background in LO results mainly from the QCD-Compton 
process (QCDC), see Fig.~1. 
For events with no hard QCD matrix element, namely
'quark parton model (QPM)' events, the $p_t$ of the hadrons 
only comes from 
fragmentation and intrinsic $k_t$ of the partons in the proton. Hence
this background is expected to be strongly suppressed if 
two particles with a 
$p_t$ above 1-2 GeV are required.

 We make a full Monte Carlo simulation of the signal and background
processes, include hadronization and higher order effects via parton
showers. Starting from three different sets of 
polarised gluon distributions, shown in Fig.~2, 
we check the sensitivity of the 
measurements and extract $\Delta G(x)$. 
These distributions are the Gehrmann-Stirling 
(GS) sets A and C~\cite{GS}, which
result from a QCD analysis of $g_1$ data, and the instanton-gluon
distribution~\cite{Kochelev}. The latter results from a calculation
of the polarised parton distribution in the Instanton Liquid 
Model~\cite{Instanton}.
The distributions differ substantially and are purposely selected to 
demonstrate how poorly $\Delta G(x)$ is constrained by 
 the present polarised data. 
For the unpolarised parton density functions the parametrizations 
of Gl\"uck, Reya and Vogt in LO were used~\cite{GRV94}.

\section {Analysis method and expected asymmetries}

\vspace{1mm}
\noindent
To estimate the sensitivity of high-$p_t$ track events to $\Delta G$
a study very similar to the one using di-jet events~\cite{gaby} was
performed.
The Monte Carlo program PEPSI~6.5~\cite{PEPSI1,PEPSI65} was
used and the kinematic range in $Q^2$ and $y$ was taken to be:
 $5<Q^2<100$~GeV$^{2}$ and $0.3 < y < 0.85$.
PEPSI includes, apart from the polarized cross sections at LO, hadronization
 and (unpolarized) parton showers for higher order effects. 
The $z-s$ scheme was used to regulate  the divergencies of the 
matrix elements~\cite{Korner,maul1}.

\begin{figure}[t]
\vspace{-4cm}
\epsfxsize=12cm
\hfil \epsffile{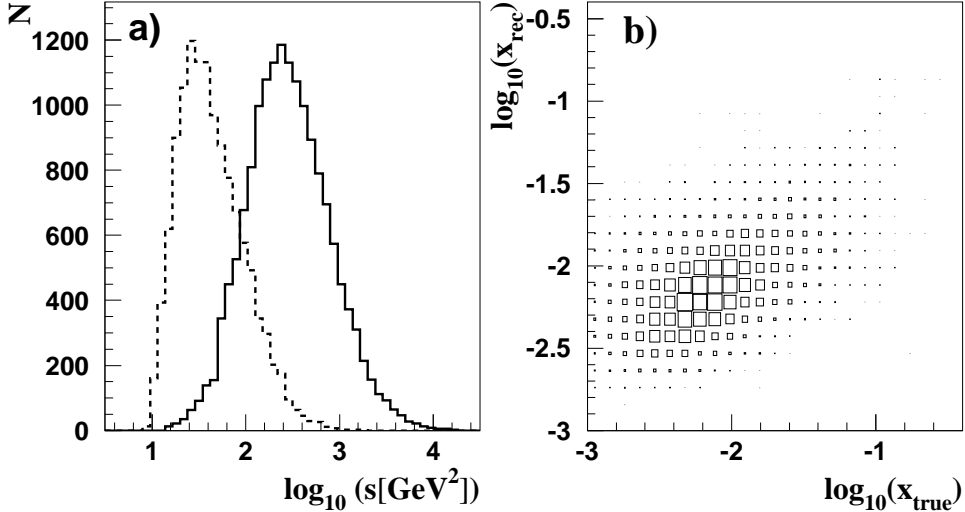} \hfil
\vspace{-5cm}
\hspace*{8cm}
\caption{a) The true $s_{ij}$ distribution of the selected 
events (full line) and the  $s_{ij}$ reconstructed  
from the two tracks (dashed line). 
b) The reconstructed $x_g$ of high $p_t$ track events versus the true $x_g$ of
the event. For the reconstructed $x_g$ a $p_t$-dependent correction 
has been applied (see text).} 
\label{fig-reco}
 \hspace*{12cm}
\end{figure}

For this analysis the acceptance of the H1 detector~\cite{H1}
 is assumed. Tracks
are detected in either  the forward or central tracking chambers. 
The acceptance for  tracks in 
 pseudorapidity $\eta$  in the laboratory frame  
amounts to the range: $-1.5 < \eta < 2.5$.
Events were selected if two tracks were found inside this $\eta$-range
 each with a $p_t > 1.5$~GeV. 
The difference in the azimuthal angle 
$\phi$ between the two tracks should be within $180 \pm 60^{\circ}$ and
the difference in pseudorapidity should be $|\eta_{track1}-\eta_{track2}| < 2$.
In case the two highest $p_t$ tracks of the event do not fulfil these 
criteria, it was checked whether a third track with $p_t > 1.5$~GeV
was present, which does fit the criteria.
In the following the convention will be that for the two selected
tracks with $p_t(1)$ and $p_t(2)$ we have: $p_t(1)>p_t(2)$.
With these cuts and the parton distributions as given above the selected 
sample contains about 85\% PGF events and 15\% QCDC events. The QPM
background is less than a percent.
Events samples with a higher 
$p_t$ cutoff were studied as well, but the gain in purity and 
kinematic variable reconstruction (see below) was outweighted by the 
loss in statistics.

\begin{figure}[t]
\vspace{-4cm}
\epsfxsize=12cm
\hfil \epsffile{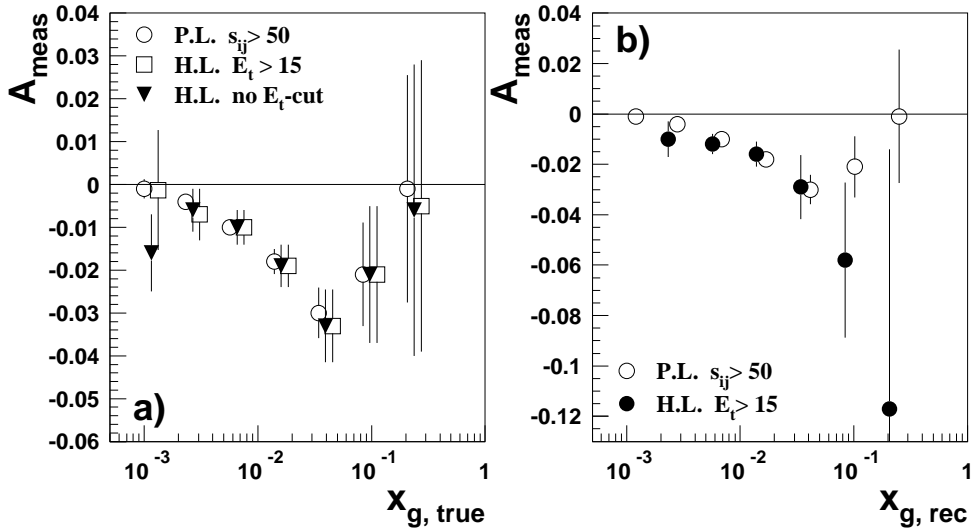} \hfil
\vspace{-5cm}
\hspace*{8cm}
\caption{a) Measured asymmetries for high-$p_t$ track events, as a function
of the true $x_g$. Asymmetries on the parton level (P.L.)
are compared with the hadron level (H.L.) for different selection cuts.
b) The measured asymmetry on hadron level versus the reconstructed
$x_g$. For comparison the same parton level points of a) are shown again.
The assumed luminosity is 200~pb$^{-1}$.}
 \hspace*{12cm}
\label{fig-asys}
\end{figure}

A
 difference compared to the di-jet analysis is the 
 reconstruction
of the kinematics of the events. In the di-jet case 
an attempt is made to reconstruct
the kinematics of the  original  parton.
Thus, the invariant mass squared of the two jets 
 is close to and well correlated with  the true
invariant mass of the hard subprocess.
Here, where only two 
tracks opposite in azimuth are considered,
 the invariant mass squared $s_{ij}^{rec}=(P_1+P_2)^2$, with $P_{1(2)}$ being
the four-momentum of  track~1(2),
 is always much
smaller than the true value of $s_{ij}$. 
This is shown in Fig.~\ref{fig-reco}a.
There is an offset between true and reconstructed $s_{ij}$,
but they are correlated.
The correlation for $x_g=x (1+s_{ij}/Q^2)$ reconstructed from
 the tracks and the true value is shown in Fig.~\ref{fig-reco}b.
Here $x$ is the Bjorken-$x$ variable, and $x_g$ is the momentum fraction
of the proton carried by the gluon for the PGF process.
The offset observed in Fig.~\ref{fig-reco}a
 due to the incomplete $s_{ij}$ reconstruction is 
already corrected for in Fig.~\ref{fig-reco}b. 
This was done by comparing the ratio of true $x_g$ 
and $x_g^{rec}$ in bins of $p_t(2)$. A polynominal
was fitted to $\log_{10}(x_g^{rec})/\log_{10}(x_{g})=a_2p_t^2(2) + a_1 p_t(2) + a_0$ and the right side of this expression was used as a multiplicative
correction factor to $x_g^{rec}$. The coefficients used were: $a_0=0.855,
a_1=-0.052, a_2=0.0018$. Values of $x_g^{rec}$ presented in this paper are
always corrected with this factor.
It can be seen that the correlation is already  rather good  with this simple
correction factor. A more sophisticated analysis, e.g.\ using a real unfolding
method, where bin-to-bin correlations are fully taken into account, can
be performed when  real data have been collected,
 and even better  results can be expected.

With the cuts detailed and used so far,  the sample contains still a
considerable amount, about 23\%, of events  with $s_{ij}<100~{\rm GeV}^2$
for which 
 NLO corrections could be potentially large (based on the experience 
with the di-jets).
An additional criterium is used   to reduce this fraction. 
One possibility is to increase the $p_t$-requirement for the tracks. 
Raising it from 1.5~GeV to 2~GeV decreases the fraction of events with
$s_{ij}<100~{\rm GeV}^2$ to about 10\%, but the total number of events is
decreased by more than 40\%.
A better possibility seems to be to check the total transverse energy
$\Sigma E_t$ deposited in the  calorimeter 
(i.e. with $|\eta| < 2.8$ in the laboratory frame) for  the event.
Requiring $\Sigma E_t > 15$~GeV reduces the fraction of low-$s_{ij}$
events  to about 10\%, but removes only about 25\% of the total statistics.
For $\Sigma E_t > 20$~GeV  a reduction of low-$s_{ij}$ events 
 to 5\%
can be obtained, while keeping 55\% of the original statistics. The latter
number is comparable to the result with the cut of $p_t>2$~GeV.
We decide here for a cut on $\Sigma E_t > 15$~GeV as a reasonable
compromise between statistics and low-$s_{ij}$ contamination.

The final event sample contains about 80,000 events/100 pb$^{-1}$.
The measurable asymmetries $A_{meas}$ for the final event sample are shown
in Fig.~\ref{fig-asys}. $A_{meas}$ is defined as in~\cite{gaby}:
\begin{equation}
A_{meas} =  \frac{N^{\uparrow \downarrow}-N^{\uparrow \uparrow}}
{N^{\uparrow \downarrow}+N^{\uparrow \uparrow}} = P_e P_p D A_{2tracks}
\end{equation}
The quantities $N^{\uparrow \downarrow}$ ($N^{\uparrow \uparrow}$)
are the total number of observed  events with two high $p_t$ tracks 
($N^{\uparrow \downarrow}=N^{\uparrow \downarrow}_{PGF}
+N^{\uparrow \downarrow}_{QCDC}$) with 
 proton and electron spin  antiparallel (parallel) to each other. 
$A_{2tracks}$ is the true physical asymmetry and on parton level the same as 
$A_{di\mbox{-}jet}$ of ref.~\cite{gaby}.
The depolarisation factor $D$ is given by $D=(y(2-y))/(y^2+2(1-y)(1+R))$,
where  $R$ is the ratio of longitudinal to
transverse $\gamma^* p$ cross section.

Figure~\ref{fig-asys}a shows the asymmetries plotted versus the true
$x_g$ of the event. The error bars correspond to an
integrated luminosity of 200~pb$^{-1}$. The assumed polarised gluon 
distribution is gluon set A of Gehrmann and Stirling~\cite{GS}.
Compared are the asymmetries on hadron level with
and without the cut of $\Sigma E_t > 15$~GeV. The
asymmetries are not changing significantly and the statistical errors 
for the low $x_g$ bins are 
slightly increased after the cut has been applied. 
Both asymmetries can also be compared to the asymmetry expected on
parton level, where  only kinematical selection cuts were applied plus a
cut on the true $s_{ij} > 50$~GeV$^2$.  The
asymmetries on parton and hadron level agree very well, the statistical
errors however increase when applying the selection on hadron level.
In order to check that the  asymmetry that can be related to 
$\Delta G$ we show in Fig.~\ref{fig-asys}b the asymmetry plotted as
a function of the reconstructed $x_g$. The parton level asymmetry is 
exactly the same as in a), i.e.~plotted versus the true $x_g$.
Hence, also for reconstructed quantities an asymmetry 
is observed, not too different in size from the one at the parton level.

The QCDC background is less than 15\% in all bins except the 
two highest $x_g$ bins where,
with increasing $x_g$, the background increases to 30\% and 50\% respectively.
The $\Delta q/q$ distributions are
however expected to be known with a precision of better than 10\% at the 
time of this measurement, and can be subtracted.
The background in the high $x_g$
region could be also further suppressed selecting hadrons with 
opposite charge or with strangeness, as demonstrated in~\cite{bravar}.


\section{Extraction of {\boldmath $\Delta G$} }
\begin{figure}[t]
\epsfxsize=9cm
\hfil \epsffile{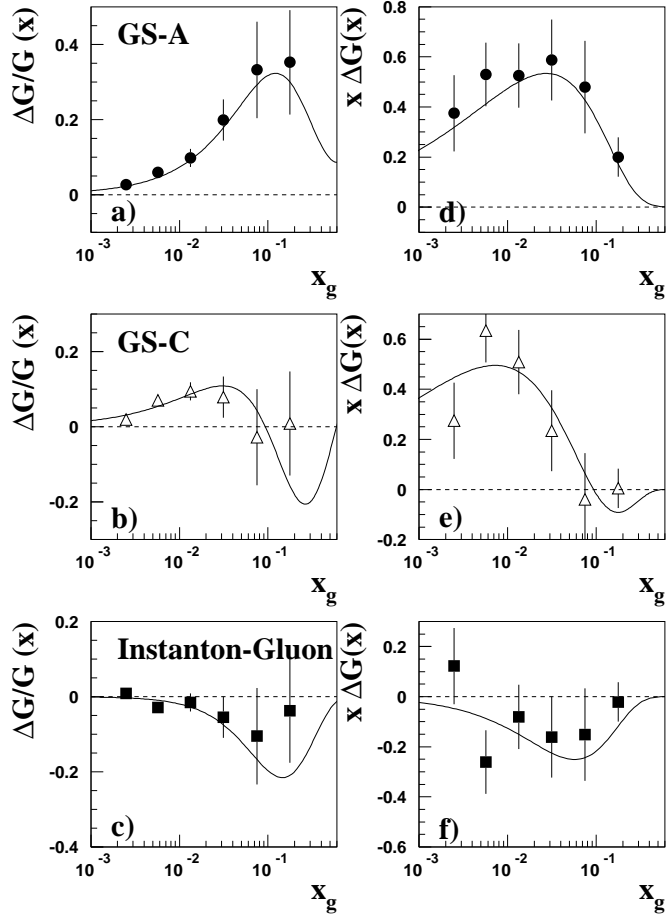} \hfil

\vspace{-1cm}
 \hspace*{12cm}
\caption{ For a luminosity of 200~pb$^{-1}$ the sensitivity to extract
$\Delta G/G$ (a-c) and $x \Delta G$ (d-f) 
is shown for different polarised gluon densities (see text).}
\label{fig-g-200}

\end{figure}

The study of the extraction of $\Delta G/G$ and $x\Delta G$
follows very closely the 
procedure described in~\cite{gaby}. 
We simulate  350~pb$^{-1}$ of events with  
GS-A as input gluon density. This  would in a real measurement
correspond to the Monte Carlo generation of events and will therefore 
be called
the 'MC-set' here.
Assuming that for each $x$-bin 
$\Delta G/G$ and the background corrected asymmetry
$A_{corr}$ are related to each other  by a simple factor $F_i$:
$$ \left(\frac{\Delta G}{G}\right)_i = F_i \cdot A_{corr,\; i},$$
where $i$ indicates the $x$-bin,
 we compute
these factors using the MC-set. ($A_{corr} = \frac{N^{\uparrow \downarrow}_{PG
F}-N^{\uparrow \uparrow}_{PGF}}
{N^{\uparrow \downarrow}+N^{\uparrow \uparrow}}$.)
The factors $F_i$ were then multiplied with the asymmetry $A_{corr}$ which
corresponds to the measured asymmetry on hadron level in Fig.~\ref{fig-asys}b.
Three 'data sets' are considered: the standard
one based on gluon set A, and   
 two additional  samples with 
 gluon set C ~\cite{GS} 
and the  instanton induced gluon density~\cite{Kochelev} respectively. 
All of these data sets correspond to a luminosity
of 200~pb$^{-1}$. They were used 
 to compute the asymmetries which were then multiplied with the same
factors $F_i$ calculated with gluon set A.
The results are shown in Figs.~\ref{fig-g-200}a - \ref{fig-g-200}c.
It can be seen here that within the statistical accuracy the input
distributions (solid lines)  
$\Delta G/G$ are properly extracted in all cases. 
 A discrimination
between different polarised gluon sets is thus possible.
Figures~\ref{fig-g-200}d - \ref{fig-g-200}f show the same result presented
for the theoretically more interesting quantity $x \Delta G$. The
error bars are scaled errors of the left column, i.e.\ no uncertainty
was assigned to $G$ which is expected to be known to better than 5\%
at the time of this measurement~\cite{klein}.

\begin{figure}[t]
\epsfxsize=9cm
\hfil \epsffile{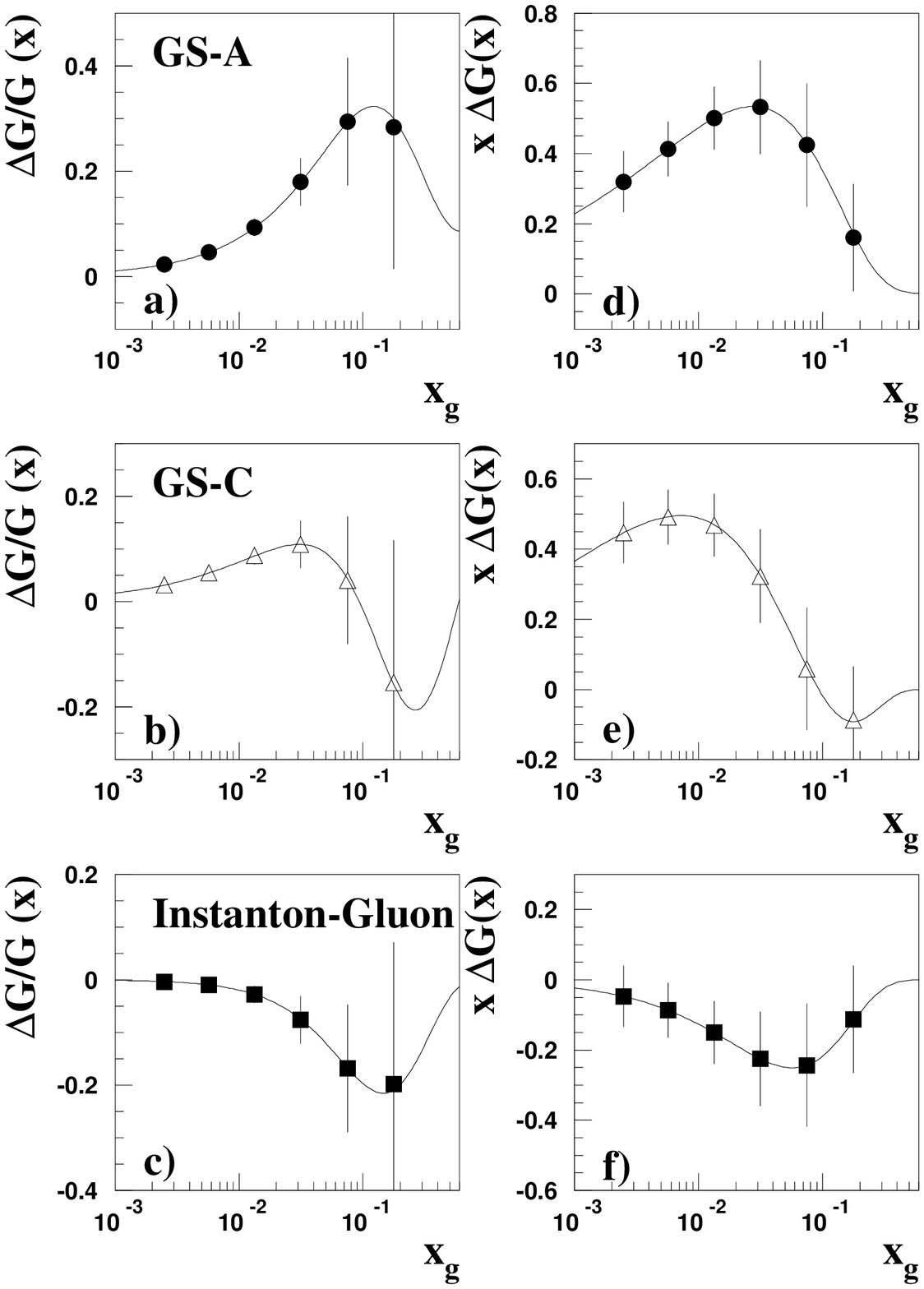} \hfil
\caption{Sensitivity to $\Delta G/G$ (a-c) and $x\Delta G$ (d-f)
 for a luminosity
of 500~pb$^{-1}$.}
\label{fig-g-500}

 \hspace*{12cm}
\end{figure}

The polarised gluon distributions used for the three 
data samples (see Fig.~\ref{fig-g-200})  differ substantially but are all
compatible with the  present polarized 
data, stressing the need for direct measurements
of $\Delta G(x)$. The GS-A and GS-C distribution show a similar small
$x$ behaviour, but differ considerably in the region around $x \sim 0.1$.
The GS-C distribution is negative for this $x$ region.
The instanton-gluon is quite different from the GS sets. It remains 
negative over the full $x$ range. The latter gluon is used in combination with 
the GS-A quark distributions for the study in this paper.

\vspace{1mm}
\noindent

The sensitivity to the shape of 
$\Delta G/G$ and $x \Delta G$ for an integrated
luminosity of 500~pb$^{-1}$
 is shown in Fig.~\ref{fig-g-500}. The statistical errors
for the  $x$ points are shown on the curves 
for $\Delta G/G$ (a-c) and $x \Delta G$ (d-f), showing the separation
power of this measurement.

The statistics is similar to the case when di-jet event are
selected~\cite{gaby}, in fact overall the track sample contains about 15\%
more events than the di-jet sample. However the events are differently
distributed in the $x_g$-bins. In this analysis we obtain more
events at  low $x_g$  than in the di-jet case but less events at very
high $x_g$. This is reflected in the statistical errors. 
The access to lower $s_{ij}$ values allows 
to explore the extraction
of the polarised gluon distribution at lower $x_g$ values, compared to the
di-jet method. Due to the presently as yet unknown higher order corrections
for the small $s_{ij}$ region, which could be potentially large, we have
not pursued this further in this study.
The rather
big error in the highest $x_g$ bin can probably reduced with a more
sophisticated reconstruction method. With the simple correction
method used in this analysis
most events originating from this bin are smeared into the
second high $x_g$ bin.

Comparing the di-jet~\cite{gaby} 
and the 2-track sample event-by-event we find that
about 40\% of the 2-track events are also selected in the di-jet analysis.
Thus a rather  large fraction of the statistics
is uncorrelated, i.e.\ we add statistically new and independent
information to the di-jet results.
In addition the systematic errors for this measurement are partially different.
This method is not sensitive to the calibration of the hadronic calorimeter
and does not depend on a jet definition.  
One of the potentially most important systematics for this measurement is 
connected with the fragmentation functions. Therefore the analysis was
repeated using the somewhat extreme
independent fragmentation scheme instead of the
LUND string fragmentation scheme~\cite{sjos}
 in PEPSI. The results were found to be in good agreement
with each other, showing that the result is not too
sensitive
to details of these functions.

\section {Conclusion}

The use of two high $p_t$ particles to extract the polarised
gluon density at a polarised HERA has been studied.
It has been found to have a similar potential to measure 
$\Delta G$ in the range $0.002 < x_g < 0.2$ as the di-jet measurement,
with the additional possibility to reach even lower $x_g$ values, if more
sophisticated unfolding techniques can be used.
At high $x_g$ the precision of the di-jet method remains superior, though.
Both methods are subject to different systematics and can therefore be
considered as complementary.

\noindent
``These Proceedings'' refers to the Proceedings of the Workshop on Physics
at HERA with Polarized Protons and Electrons.

\end{document}